\documentstyle[12pt,epsfig]{article}
\author{Diego Harari\footnote{Email: harari@cab.cnea.gov.ar}, 
Silvia Mollerach\footnote{Email: mollerach@cab.cnea.gov.ar} and 
Esteban Roulet\footnote{Email: roulet@cab.cnea.gov.ar}\\
CONICET, Centro At\'omico Bariloche, \\
Av. Bustillo 9500, Bariloche, 8400, Argentina.}

\title{Detecting filaments in the ultra-high energy\\
cosmic ray distribution}

\begin{document}

\maketitle

\begin{abstract}
We propose and test new statistical tools to study the distribution of cosmic
rays based on the use of the Minimal Spanning Tree. The method described is
particularly sensitive to filamentary structures, as those expected to arise
from strong sources of charged cosmic rays which get deflected by intervening
magnetic fields. We also test the method with data available from the 
AGASA and SUGAR surface detector arrays.   
\end{abstract}

\section{Introduction}

The search for cosmic rays (CRs) localized excess fluxes is of utmost
importance in order to identify their sources and in this way
 start to unravel the origin
of these mysterious particles. At the highest energies at which CRs are 
observed, $E > 10~{\rm
  EeV}$ (where 1~${\rm EeV}\equiv 10^{18}$~eV), 
it is expected that charged particle trajectories straighten up
and hence CR astronomy should become feasible. 

The main statistical tool that has been used to search for possible sources
among the few highest energy events observed, e.g. by AGASA or HiRes, has been
the autocorrelation function, which was applied to quantify the abundance of 
multiplets
appearing at small angular  scales (such as
the excess of doublets and triplets found by AGASA within $2.5^\circ$ and
above $40~{\rm EeV}$\cite{ta99},
 but not
confirmed by HiRes\cite{ab04}). On the other hand, when the event 
statistics is high (as for
lower energy thresholds), the search for overdensities with respect to
isotropic expectations on specific angular scales has been used to identify
possible excesses, such as those reported by AGASA \cite{ha99} and SUGAR
\cite{be00} for EeV energies near the location
of the galactic center (but not confirmed by Auger \cite{le05}). 

In this paper we propose
the use of an alternative tool, the Minimal Spanning Tree (MST), to search for
possible CR sources. This approach
 is particularly sensitive to elongated structures in the distribution of
arrival directions and is well suited for the analysis of samples of
events with statistics from few tens up to several thousands of events, and
hence should be specially useful for the search and identification of the
first candidate sources in the new generation of high statistics UHECR
experiments, such as Auger. 
 
Cosmic rays coming from a given source suffer an energy dependent deflection
due to the magnetic fields present along their trajectories. The typical
deflection produced by a regular component of the magnetic field decreases
with increasing 
energy and after traversing a distance $L$ in a magnetic field ${\bf B}$
is given by 
\begin{equation}
\delta \simeq 5^\circ \frac{10~ {\rm EeV}}{E/Z}\Big|\int_0^L 
\frac{{\rm d}{\bf x}}{\rm kpc} \times \frac{\bf B}{\mu {\rm G}}\Big|,
\end{equation}
and hence the typical strength of the regular galactic magnetic field (few
$\mu$G), which is coherent over lengths of a few kpc, can induce deflections
of order ten degrees for $E/Z =10~{\rm EeV}$, depending on the CR arrival
  direction ($Z$ is the charge of the CR).
The turbulent component of the magnetic field is expected to produce smaller
overall deflections due to the reduced scale of the field coherence length,
adding a randomly oriented deflection to the smoother change with energy 
in the source apparent position caused by the regular field
\cite{Harari:2002dy}. The combined
effect of both components is expected to generally produce elongated and
diffused images of the  cosmic ray sources. The intergalactic random magnetic
fields, which although are expected to have reduced strength ($\le 
10^{-9}{\rm G}$) can have large coherence lengths ($\sim~{\rm Mpc}$) 
and may act over long
  distances, can also further diffuse the images of point-like CR sources, but
  the size of the induced deflections is currently under debate
  \cite{si04,do03}. 

The usual statistical tools to look for small 
scale clustering in the CR angular distribution, like the autocorrelation
function mentioned before, may not be the optimal ones
 to look for the expected
filamentary-like structures. The alternative strategy that we propose
uses the MST, which
is a unique network associated to any distribution of points that connects all
the points, without forming closed loops, and having the smallest total
length. Its main advantage is that it picks out the dominant pattern of
connectedness in a manner that emphasizes the linear intrinsic
structures. Because of this, it has successfully been used to study the
filamentarity in the large scale distribution of galaxies \cite{ba85,pe95}.
We here adapt the method to the analysis of cosmic ray data and introduce
 the statistics that can optimize the search for localized excesses.

To construct the MST for a given set of points (e.g. the CR arrival 
directions in some energy range)
one possibility is just to start from any point in the set 
and find its
closest neighbor. The segment linking  this pair of points constitutes
 the first edge of the
tree. Next, one looks among the remaining points for  that one which is the 
closest to any of the points that are already in the tree, and adds the new
edge connecting these points to the tree. This procedure is repeated until
 all the points get connected.
 Then, for a set of $N$ points, the tree is formed by $N-1$ edges, and
it is unique if the distances between all the points are different. This also
means that the tree is independent from the starting point chosen.

There are various operations which can be performed on a tree in order to
enhance its sensitivity to study a particular property, the most useful being
pruning and separating. In a pruned tree branches are removed preserving the
main skeleton. In a separated tree edges exceeding some fixed cutoff (which is
usually fixed in terms of the mean edge length) are removed. This operation
tends to remove accidental linkages among independent structures, and we will
implement it in order to isolate the most relevant structures that may be
related to individual cosmic ray sources. Pruning instead is more important to
reveal a possible global structure in  the pattern of arrival directions 
and we will not implement it here.

Different statistical properties of the MST can be analyzed to pick the
different characteristics of a given distribution of points. The simplest one
is the total length of the tree. Other possibilities include the frequency
distribution of edge lengths measured in terms of the mean edge length, for
which a sample presenting localized structures shows an excess of segments
with lengths shorter than the mean and a larger tail of segments with longer
lengths  as compared to a random distribution of points \cite{ba85} (notice
that there is hence some compensation in the net effects on the average edge
length in these cases). 

In this work we adapt the MST technique to account for the effects of
the non-uniform exposure associated to different directions in the sky that is
typical in CR experiments. We do that in section~2 by introducing a tree built
on the basis of rescaled angular distances, and by so doing the results
obtained are not biased towards high exposure regions. 
We propose in section~3 the
use of a different statistics which is specially suited to look for isolated
filamentary structures superimposed to an homogeneous background.
 The main idea is to
study the distribution of the multiplicities (i.e. the number of points)
 of the disjoint pieces of the
separated tree after having cut all the edges with size greater than the
mean edge length. We also introduce in section~3 
the concept of compact multiplicity, which
allows to give a measure of how dense the
separated event groups are, what helps
 then to rank the clusters in a more meaningful
way.
 This procedure permits also to
naturally select the groups of events that are most probably associated with
the strongest  sources.

In section~4 we test the method with simulated structures superimposed to
an isotropic background, showing that these can be efficiently recovered, 
 while in section~5 we apply the technique to the available AGASA and SUGAR
 data. 

\section{The standard MST and the exposure weighted one}

In order to develop the procedure in detail and test the results for the
particular problem of the observed distribution of cosmic rays, we will use 
simulated data sets in which the partial sky coverage and the non-uniform 
exposure of the observatory is taken into account, assuming for definiteness
an experiment located at the latitude  of the Pierre
Auger Observatory ($b=-35.2^\circ$) and with maximum observed zenith angles of
$60^\circ$.

\begin{figure}[ht]
\centerline{{\epsfig{width=3.in,file=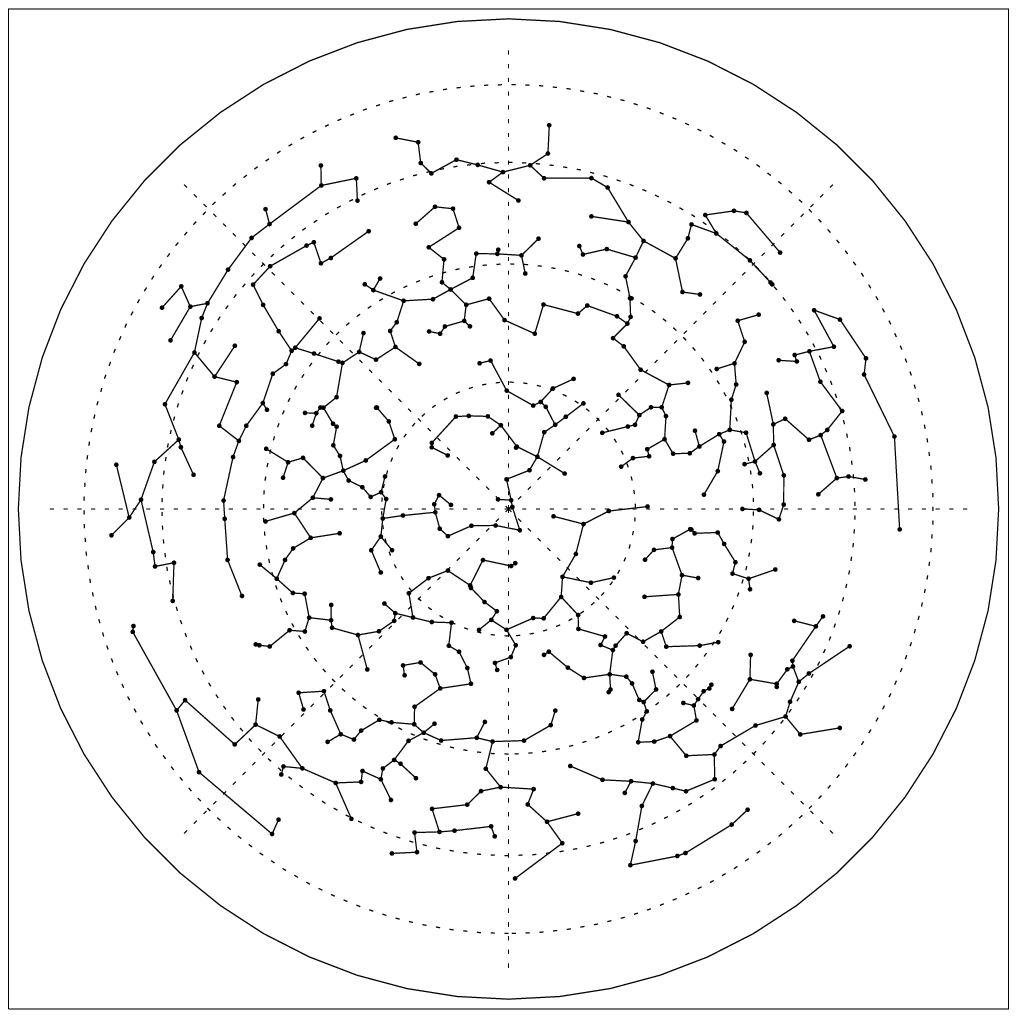}}
{\epsfig{width=3.in,file=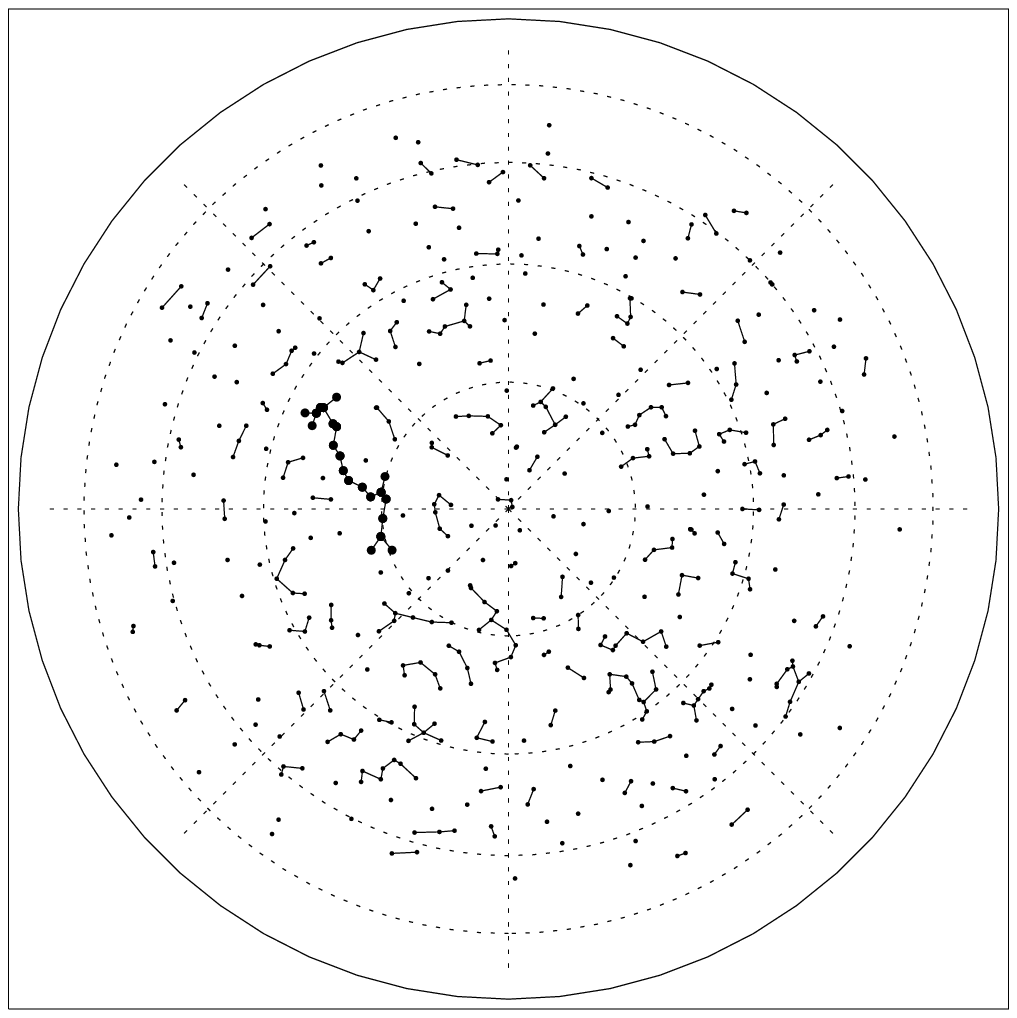}}}
\caption{Maps with 500 simulated isotropic events and the standard MSTs.
The left panel shows the whole tree while the right panel shows 
the separated one, in which the edges larger than
 the average separation among events are discarded. Big dots indicate the
 largest multiplicity object. The grid has a spacing
 of $30^\circ$ in declination.}
\label{tree.fig}
\end{figure}

In the left panel of fig.~1 we display with dots a sample of 500 random 
isotropic events
generated according to the geometric exposure corresponding to 
full acceptance for such an experiment
\cite{Sommers:2000us}.  The map is an equal area projection centered in the
southern celestial pole. 

The MST, built by joining points according to their angular distance in the sky
through the approach described before, consists of all the edges
depicted. They give a unique structure going through all points, having no
loops and minimal length. 

In the right panel of fig.~1 we display instead the separated tree, 
in which all edges larger
than the mean edge length of the original tree shown in the left panel
 are removed,
and hence only the events at distances smaller than average get associated
into clusters.
This naturally selects the regions where the density of events is enhanced,
and identifies the corresponding arrival directions.  The largest multiplicity
cluster for this simulation contains 21
events, indicated by large dots in the plot. 

One problem of this simple approach is that since the exposure depends on
declination (we are assuming for simplicity that it is uniform in right
ascension, as is approximately the case for surface detectors running for long
periods of time), the expected average distance among events also depends on
declination. Hence, having separated the tree at a common distance favors the
formation of the biggest clusters in the regions where the exposure is larger
(in this case, at declinations towards the south of the latitude
 of the detector),
and not necessarily where the strongest sources could be located.

We will hence implement a different approach by which the tree is built by
comparing not directly the actual angular distances among events, but 
instead the distances rescaled
 by the square root of the exposure (taken e.g. at the mean declination
of the two events being compared\footnote{The results are almost
  insensitive to the way in which 
this is implemented, and essentially the same is
  obtained by taking e.g. the square root of the average exposure, or the
  average of the square roots of the exposure, etc.}). In this way, when
separating the tree by the average rescaled distance, the events associated in
objects will really have neighbors at less than the average expected
separation in that direction, and no bias towards high exposure regions will
be introduced.

\begin{figure}[ht]
\centerline{{\epsfig{width=3.in,file=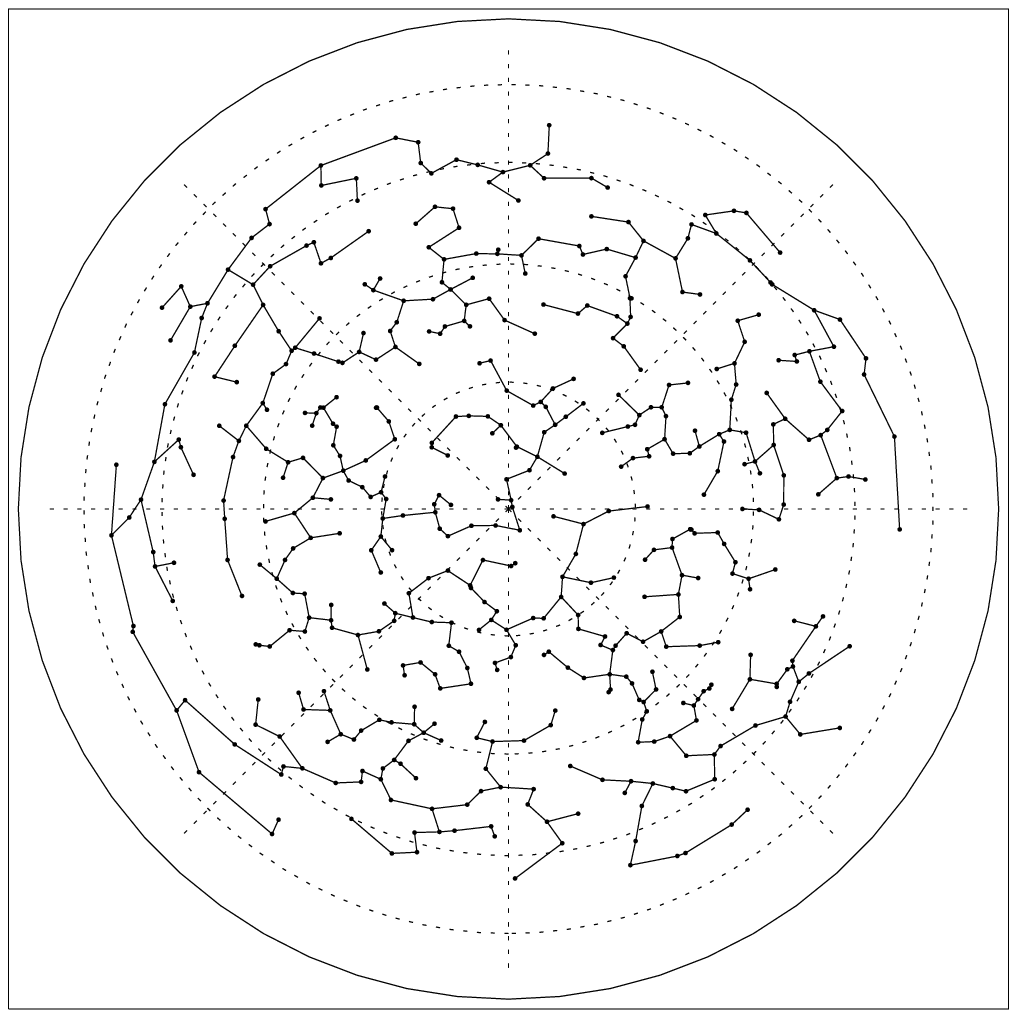}}
{\epsfig{width=3.in,file=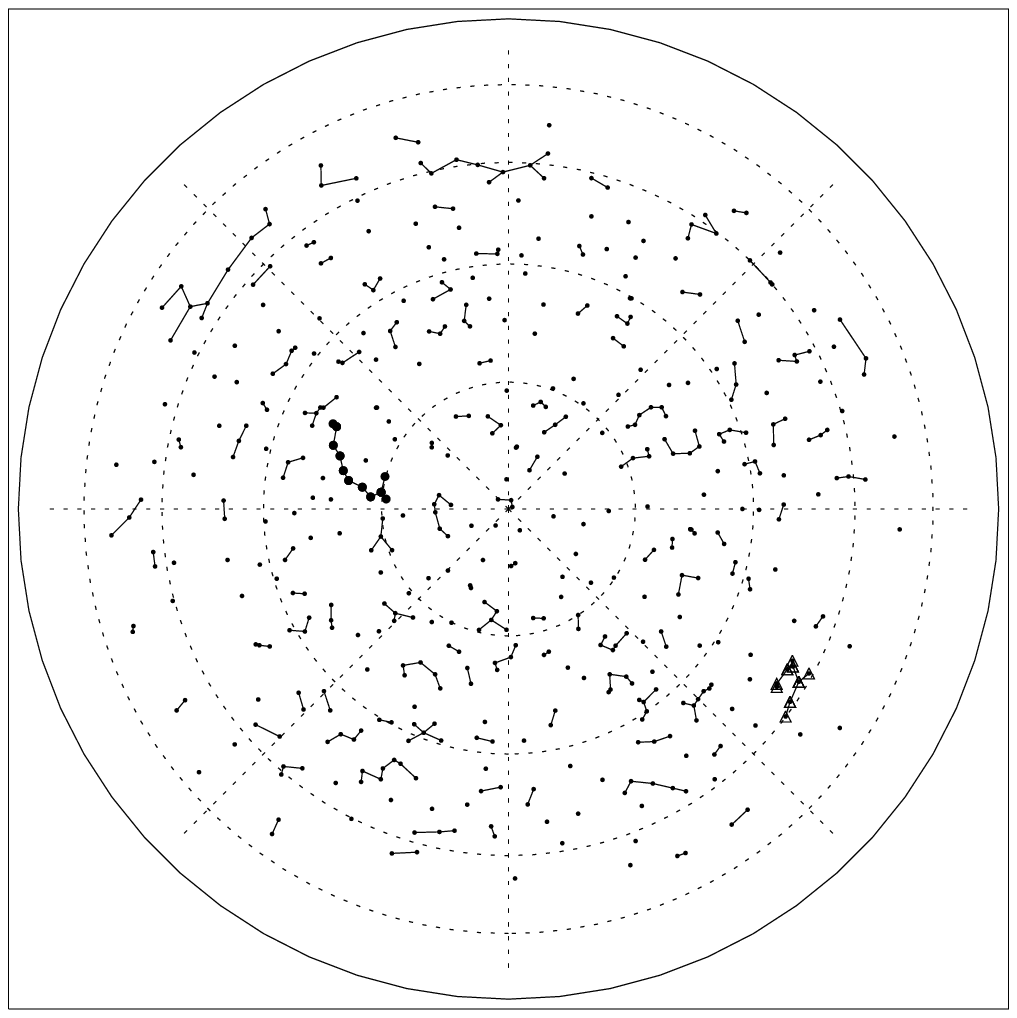}}}
\caption{Maps with 500 simulated isotropic events (the same as in
 Fig.~\ref{tree.fig}) and the exposured weighted MSTs.
On the left is the whole tree while on the right the separated one.}
\label{treecomp.fig}
\end{figure}

The left panel of fig.~2  shows the new tree just discussed with the same 
events as in fig.~1, 
while the right panel depicts the associated separated tree. 
We see that now the large multiplicity objects are distributed more uniformly
in declination (compare with fig. 1). 
The object with the
largest multiplicity, indicated with large dots, turns out now to be different.

Notice from the trees in the left panels of figs.~1 and 2 that many events near the border of
the region observed (at declination $\delta\simeq 25^\circ$) are linked 
among themselves. This
border effect is easy to understand  thinking for instance on how the tree of
a large region is affected if this region is divided into two. Many events
near the border between the two sub-regions, which before the division were
linked to an event in the other sub-region, will now have to be linked to a
different nearby event in the same subregion. Hence, many links among events
near the border result. However, these new links are usually larger than
average, and hence one doesn't find any bias towards the borders in the
separated trees of the right panels of figs.~1 and 2. In those figures there
is in addition a visual  effect associated to the polar projection used, which
tends to induce azimuthal elongations far from the South pole, but this has no
connection with the genuine effect discussed above.

\section{Standard multiplicity and compact multiplicity}

A further relevant issue for the identification of event clusters associated
to potential  strong CR sources is the compactness of the separated objects of
the tree. It is indeed clear that if two objects have similar multiplicities,
the one for which the events are closer together (using the exposure weighted
angular distance introduced in the previous section) will be the more
interesting one. Moreover, even clusters of low multiplicity may be quite
relevant if the events are sufficiently nearby, such as was the case for the
doublets and triplets observed by AGASA. We will hence introduce the concept
of {\it compact multiplicity} $\hat N_i$, which is just the multiplicity
$N_i$ of the separated tree object, divided by the average scaled angular
distance $\bar\alpha_i$ of that object 
(in units of the average $\bar\alpha$ for the whole tree before separation), 
 i.e.
\begin{equation}
\hat N_i=N_i\frac{\bar\alpha} {\bar\alpha_i},
\end{equation} 
with
\begin{equation}
\bar\alpha_i\equiv \frac{1}{N_i-1}
\sum_{j=1}^{N_i-1}\alpha_j .
\end{equation}
Here the sum is over  the $N_i-1$ edges in the object and $\alpha_j$ is the
scaled angular distance of the edge\footnote{In the case
 in which the angular extent of the object $\sum_{j=1}^{N_i-1}\alpha_j$ 
turns out be smaller  than the rescaled 
angular resolution of the experiment, $\alpha_{res}$,
we adopt $\alpha_i=\alpha_{res}/(N_i-1)$ in order not to artificially 
overweight those clusters.}
.
Notice that the average separation of the events within
objects, $\bar \alpha_i$, is always smaller than the average separation
$\bar\alpha$, and hence $\hat N_i>N_i$.

For instance, the object with the largest compact multiplicity in the
isotropic simulation of figs.~1-2 is the one indicated with triangles in
fig.~2 (right panel).
 It has a multiplicity $N_i=9$, while $\hat N_i=19.1$, since the
average separation between events in that object is $\sim 0.47$ of the global
average separation. 

\section{Identifying CR sources}

Since the autocorrelation function $C(\theta)$ counts the number of pairs with
separation smaller than a given angle  $\theta$, it is particularly sensitive
to the presence of clusters at very small angles (much smaller than the
average separation among events), for which very few pairs are expected to
result from chance coincidences. This was indeed the case for the signals reported by
AGASA above 40~EeV,  which involved several clusters 
(5 doublets and one triplet) essentially at the scale of the angular
resolution of the experiment, $2.5^\circ$, while the average separation among
the 59 events of the sample was $> 10^\circ$.

However, since the number of pairs expected grows as $\theta^4$ (for small
$\theta$),  the significance of a few 
clusters of events with separations comparable to the mean
angular separation of the events will be diluted by the large number of chance
pairs normally  expected at those angular scales\footnote{An exception 
is when a large
scale pattern, such as a dipole, is present in the data, in which case the
overdense region leads to an enhancement in the observed number of pairs at
large angular scales.}. On the contrary, the MST technique  is well suited to
identify this kind of overdense regions, since it efficiently isolates the 
clustered
events and can hence determine that they are unusual.

\begin{figure}[ht]
\centerline{{\epsfig{width=3.in,file=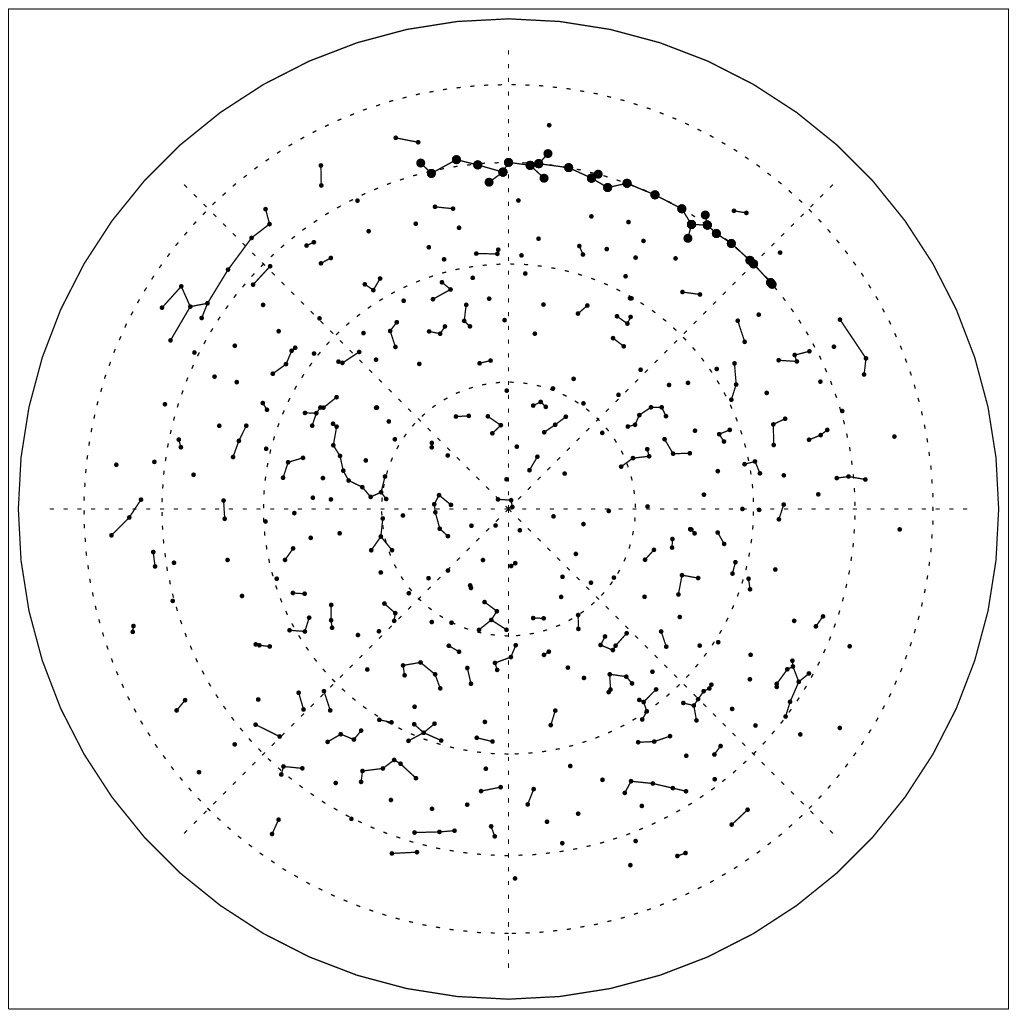}}
{\epsfig{width=3.in,angle=0,file=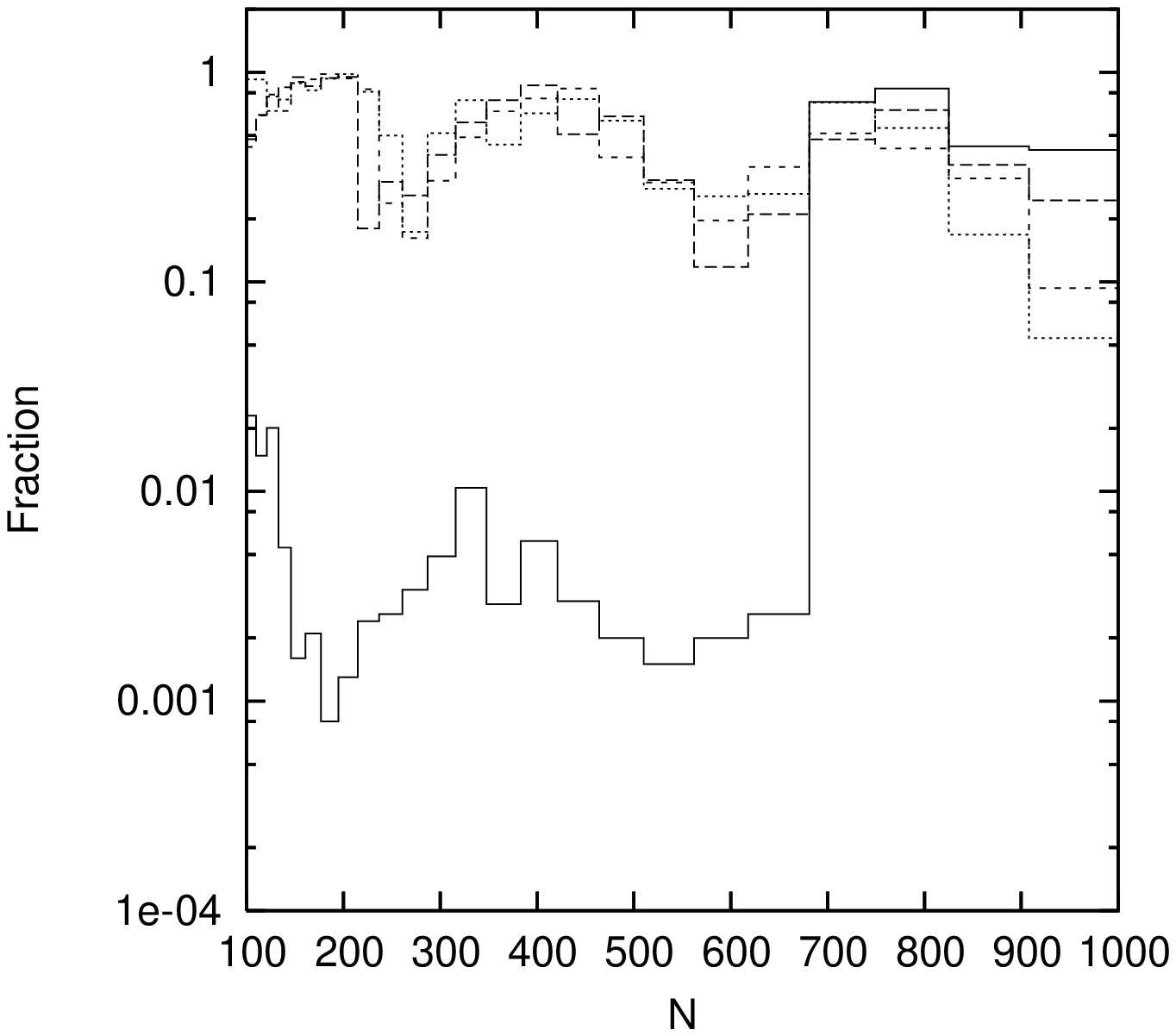}}}
\caption{Left panel: 
map with 500 events consisting of a simulated isotropic background
 with the addition of a
  string of 10 events along the equator, separated by $5^\circ$ one from the
  other. The exposure weighted separated MST is superimposed.
Right panel: Fraction of isotropic simulations having a cluster with 
compact   multiplicity larger that the data in the left panel, as a  function
of the total number of simulated events.  
Different lines correspond
 to the first, second, third and fourth largest multiplicity objects (in
 decreasing order of dash length).}
\label{source.fig}
\end{figure}

To see this we show an example in which an elongated structure consisting of
10~events aligned along the equator and separated by $5^\circ$ among each
other is superimposed to isotropic simulations (with the 
 exposure corresponding to full acceptance at the Auger latitude) 
with different total number of events $N$. 

The simulated data set  for a total of 500~events is shown in the left panel
of fig.~\ref{source.fig},  together
with  the
associated separated tree (using the rescaled angular distances). 
 The largest compact multiplicity cluster, indicated with large dots, is 
indeed associated to the
aligned events introduced by hand. In the right panel
 we show (solid line) the fraction of isotropic
simulations leading to a cluster with compact
multiplicity larger than the largest one found
in the simulated data set with the elongated source, 
as a function of the total number
of events $N$ considered.  This fraction is quite small ($< 10^{-2}$)
up to $N\sim 600$.
Notice that the average separation among events is $\sim
10^\circ/\sqrt{N/100}$, and hence we see that the fraction is particularly 
 small when the average separation among the events from the source is smaller
than the average separation among the background events.  In this example the
autocorrelation function is not sensitive to the added
source. For instance, at the scale of $5^\circ$ where the autocorrelation is
most sensitive to the structure put in by hand, already for a total of 70
events the fraction of simulations having larger number of pairs at that scale
is $>10$\%.
Similarly, a blind search of overdensities in circular windows would not have
been sensitive to the elongated structure introduced.
It has to be noticed that both the autocorrelation and the overdensity
searches, contrary to the MST, require a scan over the angular scales
analyzed, besides the scan over energy that is common to all techniques.

\begin{figure}[ht]
\centerline{{\epsfig{width=3.in,file=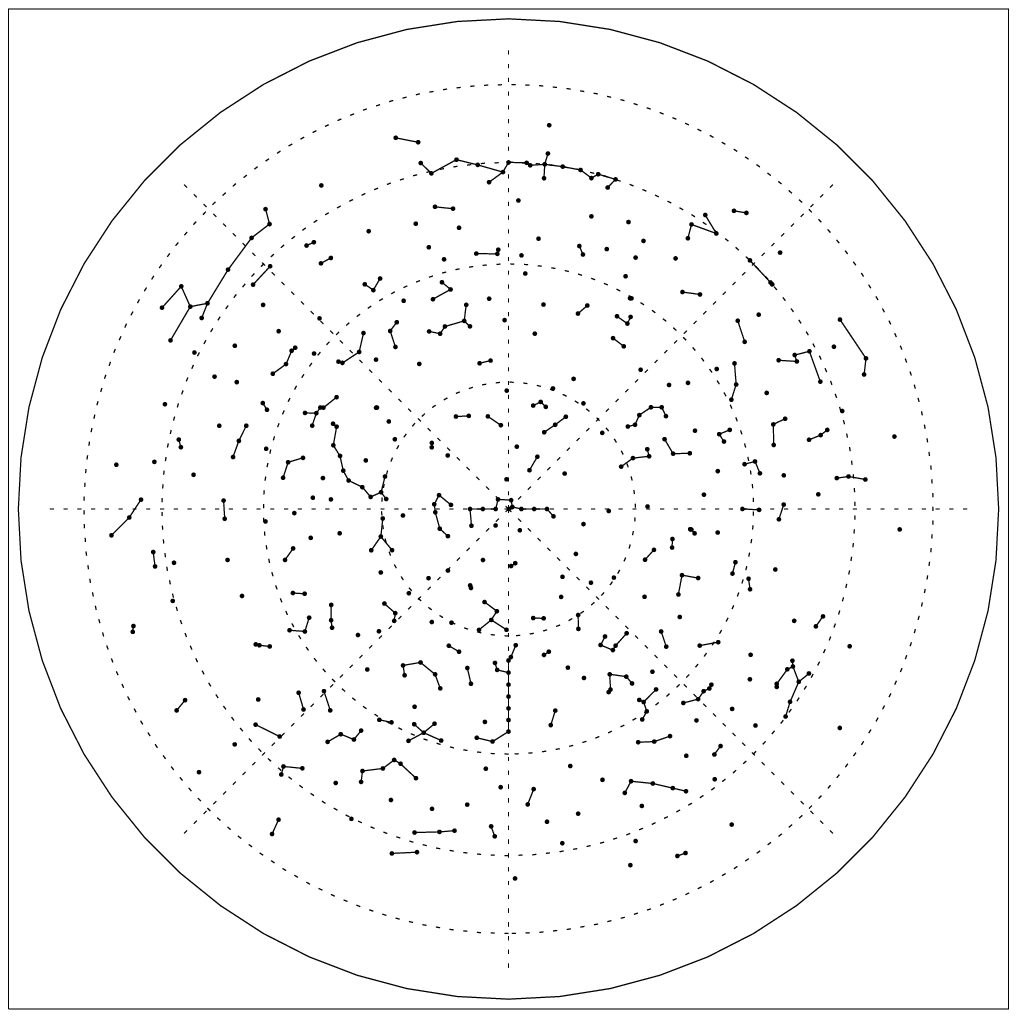}}
{\epsfig{width=3.in,file=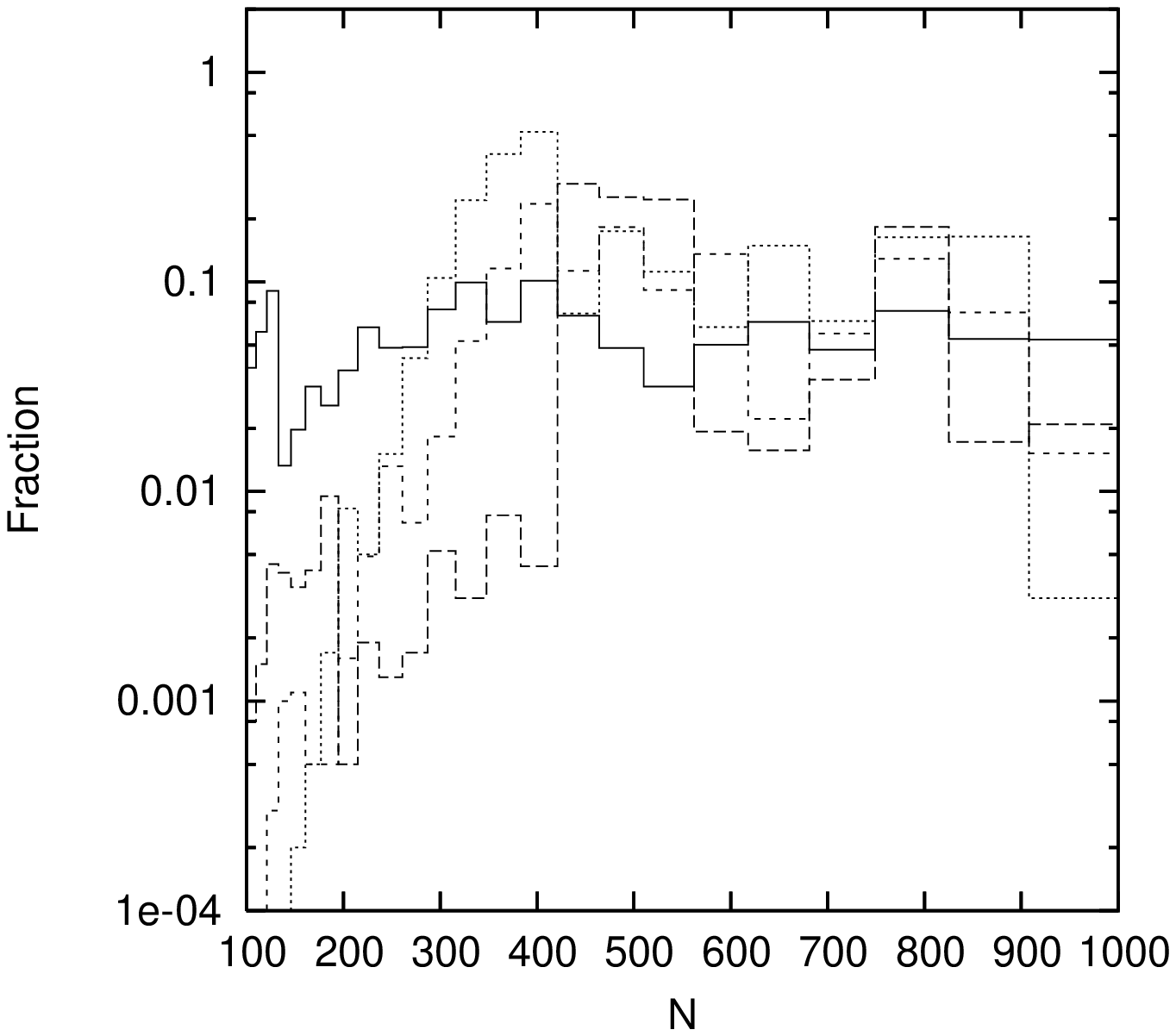}}}
\caption{
Same as fig.~\ref{source.fig}, but for 
 500 events consisting of a simulated isotropic background
 with the addition of three
  strings of 7 events, separated by $3^\circ$ one from the
  other. }
\label{src3fil73d.fig}
\end{figure}

We also show in fig.~\ref{source.fig} (right panel) 
the fraction of isotropic simulations having a
compact multiplicity  for the second largest multiplet larger than that of the
simulated data set (long dashes), and similarly for the third (medium dashes)
 and fourth (short dashes)  largest multiplets.  These kind
of tests can be useful to identify if there is more than one abnormally large
multiplet, which was not the case for the simulated data we used in this
example. 

As a second illustrative example, we show in figs.~\ref{src3fil73d.fig}
 plots similar to
those in figs. \ref{source.fig} but for the case of an isotropic simulation 
to which we
added three linear structures of 7 events each, separated by $3^\circ$ from
one event to the next. 
One of the structures is near the south pole, another
near declinations of $-45^\circ$ and the third one is along the equator.
Although the largest multiplet in the data is not very
 unlikely (since a multiplet with $\sim 7$ events is quite common for
 $N>100$), the other ones are quite unlikely up to $N\sim{\rm few}$
 hundreds. 
Notice that even if only 3 sources are put in by hand, also the
 fourth largest multiplet is unlikely, since this one turns out in general to
 be the one that would have been the largest multiplicity object in the sample
 had we not introduced the ad hoc sources.

It is important to keep in mind that once an uncommon structure is identified
by means of the MST, many detailed studies of those events can be useful. In
particular, the energy spectrum of the clustered events can be of interest,
and angle-energy correlations can provide information about the intervening
magnetic fields or about the charge composition of the CRs, they can allow to
better reconstruct the original source location, etc. \cite{Harari:2002fj}.

\section{AGASA and SUGAR MSTs}

\begin{figure}[ht]
\centerline{{\epsfig{width=3.in,file=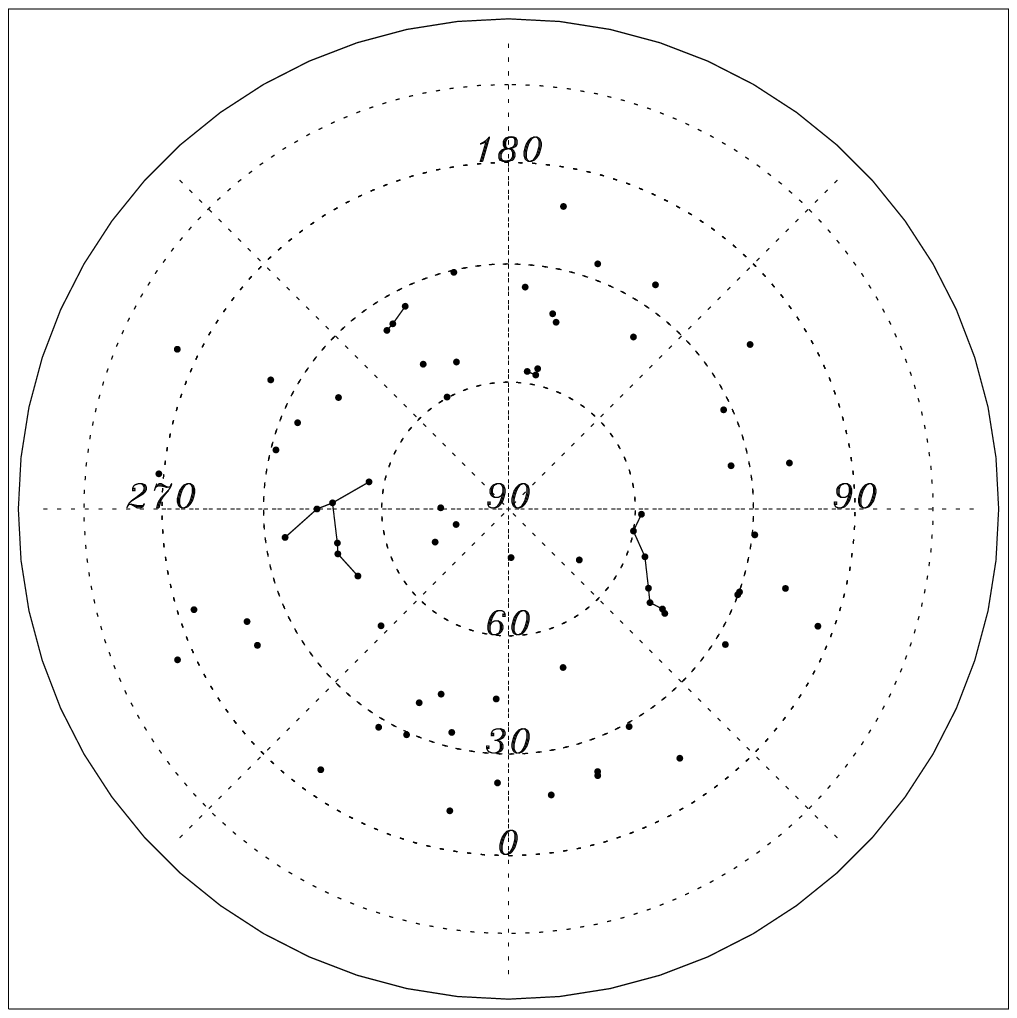}}
{\epsfig{width=3.in,file=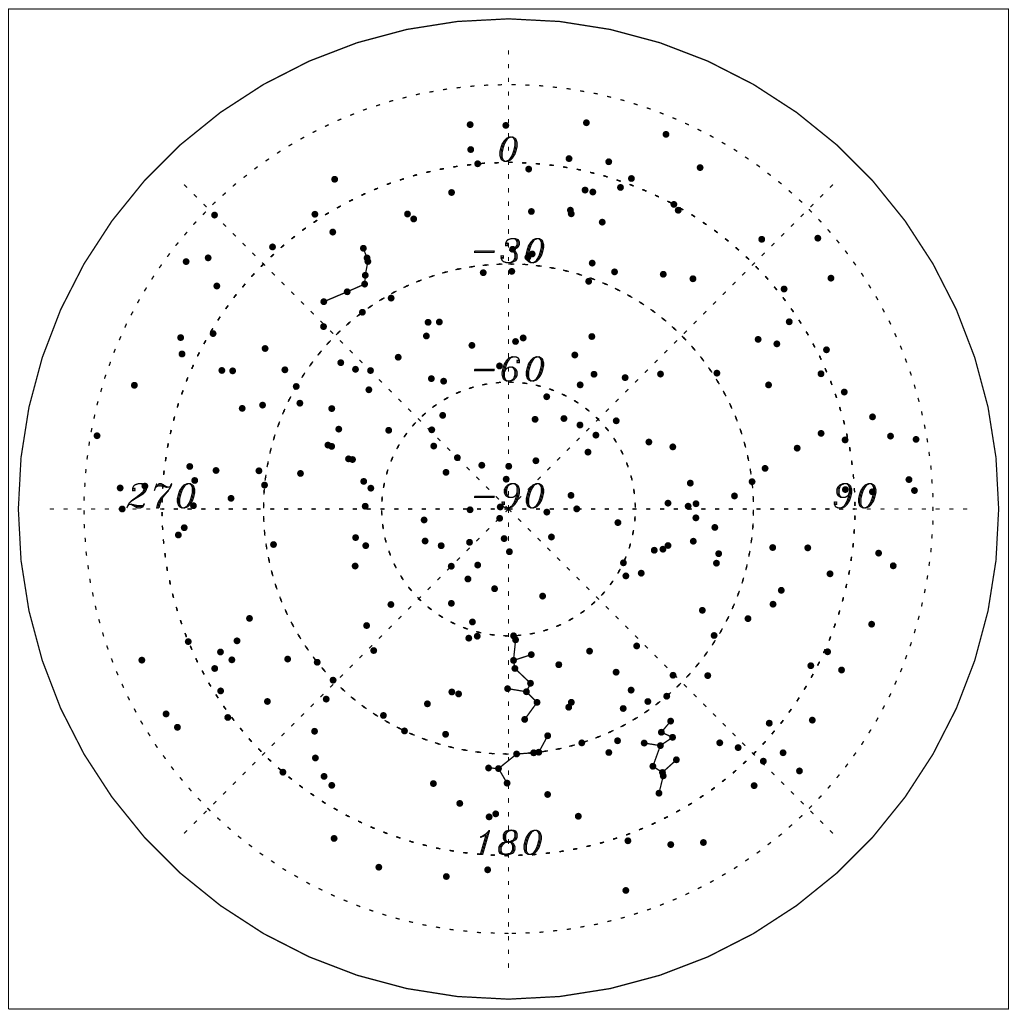}}}
\caption{Four objects with highest compact multiplicities in the 
exposure-weighted MSTs of the 72 highest energy AGASA events (left panel,
map centered in the Northern Celestial Pole) and of the 300 highest 
energy SUGAR events (right panel, map centered in the Southern Celestial 
Pole).} \label{agasa-sugar.fig}
\end{figure}

As an example of an application of the MST technique, we show in 
figs.\ref{agasa-sugar.fig} the four objects with highest compact 
multiplicities in the 
separated weighted trees obtained with the 72 most energetic AGASA events
\cite{ta99} 
(with $E > 40$~EeV and zenith angle below $45^\circ$)
as well
as the 300 more energetic SUGAR events \cite{sugar} (with 
zenith below $60^\circ$),  assuming the geometric, 
full-acceptance  exposures at the latitudes appropriate for those experiments. 
None of these objects is very unlikely to occur in random 
isotropic realizations. The fraction of isotropic simulations leading to
objects with larger compact multiplicities than the four more 
clustered AGASA objects are 22\%, 2\%, 19\% and 9\% respectively (note that
the largest compact multiplicity object found is the well known AGASA
triplet),  while
in the case of the SUGAR objects they are 50\%, 40\%, 47\% and 59\% 
respectively. No abnormally large multiplicity 
objects appear either in the MSTs 
of the SUGAR data with the 100 and 200 highest energy 
events. 

\section{Conclusions}

As a summary, we have shown that the MST provides a very 
interesting tool to
search for filamentary-like structures in the distribution of CR arrival
directions.
 We adapted the method in order to account for
the effects of the non-uniform exposure typical of CR experiments.
We introduced the study of the multiplicities of the separated clusters as an
efficient method to identify potential sources. We further
developed the concept of compact multiplicities, which allows to give larger
significance to denser concentrations of events. By comparing the largest
multiplicities observed in the data set under analysis with those obtained in
isotropic simulations with the same exposure, it is possible to characterize
how likely these configurations are and identify the events that constitute
them.  We illustrated  the power of the method with fake data consisting of
different strings of events  (representing idealized strong CR sources) 
superimposed onto an isotropic simulated background. Although the
traditionally used tools such as the autocorrelation function or the
overdensity searches would have been unsensitive to those structures, the MST
based technique introduced in this work was able to efficiently isolate the
fake sources and show that they were unusual even in the presence of a
significant number of background events.
 We then applied these tools  
to the available AGASA and SUGAR data, without finding any very significant
evidence in favor of the presence of elongated structures in those datasets. 

This method should be useful for the search of signals from powerful CR
sources in the next generation of large CR detectors, such as the Auger
Observatory, which is near completion and already gathering very large
statistics of ultra-high energy events.

\section*{Acknowledgments}
We thank support from CONICET, ANPCyT and Fundaci\'on Antorchas.

\end{document}